\documentstyle[12pt]{article}
\newcommand{\ecor}{\frac{\epsilon^{MNRS}}{V}}
\newcommand{\beq}{\begin{equation}}
\newcommand{\eeq}{\end{equation}}
\newcommand{\bea}{\begin{eqnarray}}
\newcommand{\eea}{\end{eqnarray}}
\newcommand{\half}{\frac{1}{2}}
\newcommand{\ihalf}{\frac{i}{2}}

\newcommand{\khalf}{\frac{\kappa}{2}}
\newcommand{\ikhalf}{\frac{i\kappa}{2}}

\newcommand{\up}{\frac{\partial\Upsilon}{\partial\phi}}

\newcommand{\upc}{\frac{\partial\Upsilon}{\partial{\phi}^{*}}}

\newcommand{\upp}{\frac{\partial^2\Upsilon}{\partial\phi^2}}
\newcommand{\uppc}{\frac{\partial^2\Upsilon}{\partial\phi^{*}\partial\phi}}

\newcommand{\upppc}
{\frac{\partial^3\Upsilon}{\partial\phi^{*}\partial{\phi}^2}}
\newcommand{\upppp}
{\frac{\partial^4\Upsilon}{\partial{\phi^{*}}^2\partial{\phi}^2}}

\newcommand{\ups}{\Upsilon}
\newcommand{\gs}{\Gamma_5}
\newcommand{\metric}{\sqrt{-g}}
\newcommand{\G}{\Gamma}
\newcommand{\opsi}{\overline{\Psi}}
\newcommand{\oeps}{\overline{\epsilon}}
\newcommand{\opsim}{\overline{\Psi}_{M}}
\newcommand{\oxim}{\overline{\Xi}}
\newcommand{\ola}{\overline{\Lambda}}
\newcommand{\opar}{\overline{\epsilon}}
\newcommand{\opartial}{\overline{\partial}}
\newcommand{\ze}{\overline{z}}
\newcommand{\bslash}{\not\! b}
\newcommand{\aslash}{\not\!\! A}
\newcommand{\dslash}{\not\!\! D}
\newcommand{\dslashat}{\hat{\not\!\! D}}

\newcommand{\dmu}{\partial_{\mu}}

\begin{document}
\title
{SUPERGRAVITY AND A BOGOMOL'NYI BOUND IN THREE DIMENSIONS}
\author{
Jos\'e D. Edelstein\thanks{CONICET},
Carlos N\'u\~nez and
Fidel A. Schaposnik\thanks{Investigador CICBA, Argentina}\\
Departamento de F\'\i sica, Universidad Nacional de La Plata\\
C.C. 67, (1900) La Plata\\
Argentina}
\date{}
\maketitle

\thispagestyle{empty}
\def\thepage{\protect\raisebox{0ex}{\ } La Plata-Th 95-10}
\thispagestyle{headings}
\markright{\thepage}

\begin{abstract}
We discuss the $2+1$ dimensional Abelian Higgs model coupled to
$N=2$ supergravity. We construct the  supercharge algebra
and, from it, we show that
the mass of classical static
solutions is bounded from below by the topological charge.
As it happens in the global case,  half of the supersymmetry is broken
when the bound
is attained and  Bogomol'nyi
equations, resulting from the unbroken supersymmetry, hold.
These equations, which correspond to gravitating vortices, include
a first order self-duality equation whose integrability condition
reproduces the Einstein equation.
\end{abstract}

\newpage
\pagenumbering{arabic}
\setcounter{page}{1}

\section{Introduction}
The relevance of the solutions to classical equations of motion
of non-linear field theories (solitons, instantons) is nowadays
recognized both in Mathematics and in Physics \cite{raj}.

Since the pioneering work by Belavin, Polyakov, Schwartz and
Tyupkin
\cite{BPST} an important feature of many of these (second order)
equations of motion was discovered: through the obtention of a
bound of topological origin (for the energy or the action,
depending on the case), solutions can be found studying much
simpler first order
differential equations (self-duality equations or Bogomol'nyi
equations \cite{Bogo}) An increasing number of works have been
addressed to this issue in the last 20 years.

Already in ref.\cite{dVS}, where the Bogomol'nyi equations were
first written for the Abelian Higgs model, it was recognized that
the existence of these first-order equations was connected with
the necessary conditions for supersymmetry in models with
gauge symmetry breaking \cite{SSF}.

Other works then stressed this fact \cite{dVF} but the main advance
in the understanding of this question was achieved by Olive and
Witten \cite{WO} in their work on the connection between
topological quantum numbers and the central charge of extended
supersymmetry. An important result of this investigation was to show
that the classical aproximation to the mass spectrum
is exact at the quantum level since supersymmetry ensures that there
are no quantum corrections. In other words, the Bogomol'nyi's bound
is valid quantum mechanically and is saturated.

Many  models were studied afterwards following this line
\cite{HS1}-\cite{ENS}. Concerning gauge theories with symmetry breaking
(the case of interest in the present work)
the interplay between Bogomol'nyi equations and supersymmetry can
be understood as follows \cite{ENS}:
\vspace{3mm}

\noindent{For gauge theories with symmetry breaking having a
topological charge and an $N=1$ supersymmetric version, the $N=2$
supersymmetric extension, which requires certain conditions on
coupling constants, has a central charge coinciding with the topological
charge. This relation ensures the existence, of a Bogomol'nyi bound
and, consequently, of Bogomol'nyi equations}.
\vspace{3mm}

It is important to note that in the soliton sector, half of the
supersymmetries of the theory are broken \cite{hull}.

Once the connection between global supersymmetry and Bogomol'nyi
bounds was understood, the natural question to pose is whether
similar phenomena take place for {\it local}
supersymmetry including gravity.
That is, to investigate
the possibility of establishing a connection
between supergravity and Bogomol'nyi bounds for gravitating solitons
and from this, to obtain first-order differential equations whose
solutions also solve Einstein and Maxwell (or Yang-Mills) equations together.

The works on this issue are based on those of Teitelboim \cite{T},
Deser and Teitelboim \cite{DT}, Grisaru \cite{G} and Witten-Nester-Israel
\cite{W}-\cite{NI} on positivity
of the energy using supergravity. On this line, the Einstein-Maxwell theory
was investigated by Gibbons and Hull \cite{GH} and the analog of Bogomol'nyi
bounds were found for the (ADM) mass \cite{Gibb}. In the same vein,
Kallosh et al \cite{KLOPvP}, Gibbons et al \cite{GKLTT}
and Cveti\v{c} et al \cite{CGR} studied
different gravity models.

Extending our previous work on the globally supersymmetric
case \cite{ENS}, we study
in the present paper the Abelian Higgs model coupled to supergravity in
$2+1$ dimensions. We have chosen such a model in view of
the experience accumulated in the study
of Bogomol'nyi equations for vortex configurations, the connection
we have already established for global supersymmetry
and the simplicity one should
expect from a $2+1$ abelian model.
To our knowledge, there is not much work on this
model except for a recent paper by Becker, Becker and Strominger
\cite{BBS},
reported while the present investigation was in progress.
Although some of our results overlap those in \cite{BBS},
we think it is worthwhile
to present a detailed discussion of our approach
which is based,
as in the global case \cite{ENS}, in the study of the
supersymmetry algebra and provides a systematic
way of exploiting supersymmetry in the search of
Bogomol'nyi equations. Not surprisingly, being the supersymmetry
local, we find
that a  $1$ form, analogous to the Witten-Nester-Israel form used
in the proof of positivity of the energy in gravitation
\cite{W}-\cite{NI}, plays a
central role in our derivation.

Indeed, we will show explicitely that the supercharge algebra can be
realized in
terms of the circulation, over a space-like surface contour, of a
generalized Witten-Nester-Israel form adapted to the present $d=2+1$ case.
This fact will be shown to be at the root of the existence of a Bogomol'nyi
bound. Our procedure will be systematic in the sense that its formulation
is adapted
to any supergravity model where conserved supercharges could be
defined and with a bosonic sector admiting topological charges.

The plan of the paper is as follows: in Section
2 we carefully discuss the $N=1$, $d=3+1$ supergravity action
establishing our notation and conventions so that the dimensionally
reduced $2+1$ Abelian Higgs model coupled to supergravity can
be easily obtained (Section 3). Section 4 addresses to the
supersymmetry algebra and its connection with the Bogomol'nyi
bound (from which Bogomol'nyi equations can be obtained).
We give a discussion of our results in Section 5.


\section{The $N=1$ Action in $d=4$}
We shall construct the $N=2$ locally
supersymmetric Abelian Higgs model in 3-dimensional space
by dimensional reduction \cite{CJSS} of an appropriate
$N = 1$, $d = 4$ supergravity model.
To our knowledge,
it was not until very recently that the (Bosonic sector of the)
corresponding Lagrangian has been written \cite{BBS}.
This section is
addressed to the description of the Abelian Higgs model coupled to $N=1$
four-dimensional supergravity leaving for Section 3 the dimensional reduction.

Let us consider the following locally supersymmetric and
gauge invariant superspace action
for matter interacting with gravity and electromagnetism
\begin{equation}
S = \int d^4x d^4\theta E \left[
\half
\Delta (\Phi,\overline{\Phi}e^{2\tilde{q}{\cal V}})
\exp{(-\xi\kappa^2{\cal V})} + Re\left(\frac{2}{R}\overline{W}W\right)
\right]
\label{action1}
\end{equation}
Here, $\Phi$ is a chiral (matter) multiplet whose component fields are
the Higgs field $\phi$, a Higgsino $\Xi$ and an auxilliary
field $F$. ${\cal V}$ is a vector (gauge) multiplet which in the Wess-Zumino
gauge contains the photon $A_M$, the photino $\Lambda$
and a real auxilliary field $D$.
$W$ is the supercovariant strength multiplet containing the vector
field strength. The superspace determinant is denoted by $E$ and
$R$ is a chiral scalar curvature superfield \cite{WB}.
We will distinguish curved ($M,N,R,\ldots$)
and flat ($A,B, C,\ldots$) indices.
The $N=1$ supergravity multiplet contains the vierbein $V_M^A$,
the gravitino $\Psi_M$, a complex scalar auxilliary
field $U$ and a pseudovector auxilliary field $b_M$.
$\Delta (\Phi,\overline{\Phi}e^{2\tilde{q}{\cal V}})$
is an arbitrary gauge invariant functional while $\tilde{q}$ is the
$U(1)$ charge. $\xi$ is a real parameter and $\kappa$ is the
gravitational constant.
We have not included a superpotential term. The interaction
between chiral and vector multiplets
is the local version of the Fayet-Iliopoulos term \cite{FGKvP}.
We shall adopt hereafter the Wess-Zumino gauge.

After some calculations, the $N=1$ supergravity Lagrangian
(\ref{action1}) can be written in terms of component fields as
\beq
{\cal L} = {\cal L}_{Bos} + {\cal L}_{Fer}
\label{lag1}
\eeq
where
\bea
V^{-1}{\cal L}_{Bos} & = & \frac{1}{6}\Upsilon{\cal R}
- \frac{1}{4}g^{MR}g^{NS}F_{MN}F_{RS}
- \uppc(D_M\phi)(D^M\phi)^{*} \nonumber \\
& + & \frac{\kappa}{3}
b^M\left(i\up D_M\phi  - i\upc (D_M\phi )^{*} - \xi\kappa^2\ups A_M \right)
\nonumber \\
& + & \frac{\xi\kappa^2}{2}A^M\left(i\up D_M\phi
- i\upc (D_M\phi )^{*} - \frac{\xi\kappa^2}{2}\ups A_M \right) \nonumber \\
& + & \frac{\kappa}{3}\left( U^{*}\up F + U\upc F^{*}\right)
+ \uppc\vert F\vert^2 \nonumber \\
& + & \frac{\kappa^2}{9}\ups (|U|^2 - b_Mb^M)
+ \tilde{q}D\phi\up - \frac{\xi\kappa^2}{2}\ups D
+ \half D^2
\label{lagbos}
\eea
and
\bea
V^{-1}{\cal L}_{Fer} & = &
\frac{\kappa^2}{12}\Upsilon\frac{\epsilon^{MNRS}}{V}
{\opsim}\Gamma_5\Gamma_N\left({\cal D}_R +
i\frac{3\xi\kappa^2}{4}A_R\G_5\right)\Psi_S
\nonumber \\
& - & \uppc \oxim_-\left(\dslash + i\frac{\kappa}{6}\bslash
- i\frac{\xi\kappa^2}{2}\aslash\right)\Xi_+
- \frac{1}{4}\ola\left(\dslashat + i\frac{\kappa}{2}\bslash\G_5\right)\Lambda
\nonumber \\
& + & \upppc\oxim_-\dslash\phi\Xi_+ + \half\upppp\oxim_-\Xi_-\oxim_+\Xi_+
- \upppc \oxim_-\Xi_- F^{*}\nonumber \\
& - & \frac{i\kappa^2}{3}b^M\up{\opsim}_-\Xi_-
+ \frac{i}{8}\xi\kappa^3\ups\overline{\Psi}\cdot\G\G_5\Lambda
- \frac{\kappa}{3}U^{*}\upp\oxim_-\Xi_- \nonumber \\
& + & \kappa\uppc{\opsim}_-(\dslash\phi)^{*}\G_M\Xi_-
- \frac{4\kappa}{3}\up\oxim_-\Sigma^{MN}{\cal D}_M{\Psi_N}_- \nonumber \\
& - & \frac{\kappa^2}{8}\ecor{\opsim}\G_N\Psi_R \left(\up D_S\phi
+ \half\uppc\oxim_+\G_S\Xi_-\right)  \nonumber \\
& + & \frac{\kappa^3}{6}\up\oxim_-\left({\Psi_M}_- \overline{\Psi}\cdot\G
\Psi^M
+  \Sigma^{MN}\left(
\half{\Psi^R}_-\overline{\Psi}_N\G_R\Psi_M \right. \right. \nonumber \\
& + & \left. \left. {\Psi_N}_-{\opsim} \G\cdot\Psi\right)\right)
+ \ihalf\tilde{q}\kappa\up\phi{\overline{\Psi}_-}\cdot\G \Lambda_+ -
2i\tilde{q}\uppc\phi\ola_+\Xi_+ \nonumber \\
& - & \frac{\kappa^2}{2}\uppc{\opsim}_-\Xi_-{\overline{\Psi}^M}_+\Xi_+
- \frac{\xi\kappa^2}{2}\upc\oxim_+\Lambda_+ \nonumber \\
& + & \frac{\kappa}{8}\overline{\Psi}_T\Sigma^{MN}\Gamma^T\Lambda F_{MN}
+ h.c.
\label{lagfer}
\eea
Here $V$ is the
determinant of the vierbein,
\[      V(x) \equiv E(x,\theta,\overline{\theta})\vert_{\theta =
\overline{\theta} = 0} ,      \]
and ${\cal R}$ is the Ricci scalar
(depending on the vierbein $V_M^A$ and the spinorial connection
$\Omega_{MAB}$). $\ups$ is a functional of the Higgs field defined
as:
\[ \ups (\phi,\phi^{*}) \equiv
\Delta (\Phi,\overline{\Phi}e^{2\tilde{q}{\cal V}})\vert_{\theta =
\overline{\theta} = 0} ,      \]
subindices ``$-$'' and ``$+$''
indicate  left or right-handed fermions and
the ``hatted'' derivative acting on the photino $\Lambda$ is defined as
\beq
\hat{D}_M\Lambda = {\cal D}_M\Lambda +
\frac{\kappa}{4}\left(\partial_RA_S -
\frac{\kappa}{2}\overline{\Psi}_R\G_S\Lambda - (R\leftrightarrow S)\right)
\Sigma^{RS}\Psi_M
\label{hatder}
\eeq
where ${\cal D}_M$ is the covariant derivative over fermionic fields:
\beq
{\cal D}_{M} = \partial_{M} + \frac{1}{4}\Omega_{MAB}\Sigma^{AB}
\label{dercov}
\eeq
Finally,
\[ \dslash\Xi_{\pm} = \not\!\!{\cal D}\Xi_{\pm} - i\tilde{q}\aslash\Xi_{\pm}\]
is the Maxwell and general coordinate covariant derivative acting on $\Xi$.
Let us mention that a Weyl scaling of vierbein and fermion fields and a
particular choice of the (up to now arbitrary) functional $\ups$ is
necessary in order to bring (\ref{lag1}) to its canonical form.

Lagrangian (\ref{lag1}) is
invariant under the following set of local supersymmetry
transformations (with anticommuting parameter $\epsilon (x)$):
\[ \delta A_M = - \half \overline{\epsilon}_-\G_M\Lambda_+ - \half
\oeps_+\G_M\Lambda_-  \;\;\; , \;\;\; \delta V_M^A =
\khalf\oeps_-\G^A{\Psi_M}_+ + \khalf\oeps_+\G^A{\Psi_M}_-  \]
\[ \delta D = \half\oeps_- \left(\hat{\dslash} -
\ikhalf\bslash\right)\Lambda_+ - \ihalf \oeps_+ \left(\hat{\dslash}
+ \ikhalf\bslash\right) \Lambda_- \;\;\; , \;\;\;
\delta U = \half\oeps_+\G\cdot\hat{R}_-   \]
\[ \delta F = \oeps_+(\hat{\dslash} - \ikhalf\bslash)\Xi_- -
\frac{\kappa}{3}\oxim_-(U\epsilon_- + i\bslash\epsilon_+)
+ i\tilde{q}\oeps_+\Lambda\phi \;\;\; , \;\;\;
\delta \phi = \oeps_-\Xi_-   \]
\beq
\delta b_M = \frac{3i}{4}\oeps_-\left(\hat{R}_{M-}
- \frac{1}{3}\G_M\G\cdot\hat{R}_-\right) + h.c.
\label{transfsusy}
\eeq
\[ \delta \Lambda_+ = \half \left(\partial_MA_N -
\frac{\kappa}{2}{\opsim}\G_N\Lambda - (M\leftrightarrow
N)\right)\Sigma^{MN}\epsilon_+ - \ihalf D \epsilon_+ \]
\[ \delta \Xi_- = \half\left(\dslash\phi - \kappa\G^M{\opsi_M}_-\Xi_-\right)
\epsilon_+ + \half F\epsilon_- \]
\[ \delta {\Psi_M}_- = \frac{2}{\kappa}\left({\cal D}_M + \ikhalf
b_M\right)\epsilon_- +
\frac{1}{6}\G_M(U^{*}\epsilon_+ - i\bslash\epsilon_-) \]
where $\hat{R}_{M-}$ is given by:
\[   \hat{R}_{M-} = i\frac{\epsilon^{MNRT}}{V}\Gamma_5\Gamma_N
\left[{\cal D}_R + \ikhalf\gs b_R +
\frac{\kappa}{6}\G_R(U - i\bslash\gs )\right]\Psi_T  \]
and
\beq
\hat{D}_M\Xi_{\pm} = {\cal D}_M\Xi_{\pm} - i\tilde{q}A_M\Xi_{\pm}
\pm \frac{i\kappa}{2}b_M\Xi_{\pm} - \khalf\left((\hat{\dslash}\phi){\Psi_M}_-
+ F{\Psi_M}_+\right)
\eeq

We eliminate auxiliary fields and compute on-shell quantities,
the fermion Lagrangian (\ref{lagfer}) becomes:
\bea
V^{-1}{\cal L}_{Fer} & = &
\frac{\kappa^2}{6}\Upsilon\frac{\epsilon^{MNRS}}{V}
{\opsim}\Gamma_5\Gamma_N\left({\cal D}_R +
i\frac{3\xi\kappa^2}{4}A_R\G_5\right)\Psi_S \nonumber \\
& - & \half\ola\left(\dslashat - i\frac{3\xi\kappa^2}{4}\aslash\G_5\right)
\Lambda - \frac{4\kappa}{3}\left(\up\oxim_-\Sigma^{MN}{\cal D}_M{\Psi_N}_-
+ h.c.\right)  \nonumber\\
& - & \uppc \oxim\left(\not\!\!{\cal D}
- i\left(\tilde{q}+\frac{3\xi\kappa^2}{4}\right)\aslash\right)\Xi
+ V^{-1}{\cal L}_{Fer}^{int}
\label{lagferkin}
\eea
where ${\cal L}_{Fer}^{int}$ contains complicated interactions involving
fermion fields whose explicit form is not of interest for us.
It is interesting to stress the
occurrence of a charge $3\xi\kappa^2/4$ for all
fermion fields (originated in the Fayet-Iliopoulos term).

We shall often have to put all fermion fields to zero.
Given a functional ${Q}$ depending both on bosonic and fermionic fields,
it will then be convenient to define ${Q}\vert$ for
\beq
{Q}\vert \equiv {Q}\vert_{\Psi_{A},\Lambda,\Xi=0} .
\label{spb}
\eeq
Under condition (\ref{spb}) the only surviving supersymmetric transformations
are those corresponding to fermionic fields:
\[ \delta \Lambda_+\vert = \half F_{MN}\Sigma^{MN}\epsilon_+ - \ihalf
\left(\frac{\xi\kappa^2}{2}\ups - \tilde{q}\phi\up\right)\epsilon_+ \;\;\;\;\;
, \;\;\;\;\; \delta \Xi_-\vert = \half (\dslash\phi ) \epsilon_+ \]
\[ \delta {\Psi_M}_-\vert = \frac{2}{\kappa}\left({\cal D}_M +
\frac{3i}{4\Upsilon}\left[i\up(D_M\phi) - i\upc(D_M\phi)^{*}
- \xi\kappa^2\ups A_M \right] \right) \epsilon_- \]
\beq
- \frac{i}{4\kappa\Upsilon}\G_M\left[i\up(\dslash\phi) -
i\upc(\dslash\phi)^{*} - \xi\kappa^2\ups\aslash\right] \epsilon_-
\label{trafpbb}
\eeq
Concerning the Lagrangian, it becomes
\bea
V^{-1}{\cal L}\vert & = & \frac{1}{6}\Upsilon{\cal R}
- \uppc(D_M\phi)(D^M\phi)^{*} - \frac{1}{4}g^{MR}g^{NS}F_{MN}F_{RS} \nonumber
\\
& + & \frac{1}{4\ups}
\left(i\up D_M\phi  - i\upc (D_M\phi )^{*} - \xi\kappa^2\ups A_M\right)^2
\nonumber \\
& - & \half\left(\tilde{q}\phi\up - \half\xi\kappa^2\ups\right)^2
\label{lagvert}
\eea

In order to bring (\ref{lagvert}) to its canonical way one can perform
a Weyl rescaling on the Vierbein
\beq
V_M^A \rightarrow
V_M^A \left(-\frac{3}{\kappa^2\ups}\right)^{1/2} \equiv
V_M^A e^{-\frac{1}{6}{\cal J}(\phi,\phi^{*})}
\label{weyl}
\eeq
\beq
g^{MN} \rightarrow g^{MN}\left(-\frac{\kappa^2\ups}{3}\right) \;\;\;\;\; ,
\;\;\;\;\; V \rightarrow \frac{9}{\kappa^4\ups^2}V
\label{weyl2}
\eeq
where
\beq
{\cal J}(\phi,\phi^{*}) \equiv 3 \log
\left(-\frac{\kappa^2\Upsilon}{3}\right)
\label{kahler}
\eeq
If we now choose
$2{\cal J}(\phi,\phi^{*}) = -\kappa^2\phi\phi^{*}$ and redefine
$\xi = - \tilde{q}v^2/3$, the Lagrangian takes the form
\bea
V^{-1}L\vert & = & -\frac{1}{2\kappa^2}{\cal R}
- \frac{1}{4}g^{MR}g^{NS}F_{MN}F_{RS}
- \half(D_M\phi)(D^M\phi)^{*} \nonumber \\
& - &  \frac{\tilde{q}^2}{8}(|\phi|^2 - v^2)^2
\label{lagbosult}
\eea
which is the expected Abelian Higgs model Lagrangian
minimally coupled to gravity.
Note that the coupling constant of the Higgs potential is related to
the electric charge by the well-known condition
\beq
\lambda = \frac{\tilde{q}^2}{8}
\label{condition}
\eeq
This condition was originally found for the globally supersymmetric model
\cite{SSF}. As explained in \cite{ENS}
it gives a necessary
condition for extending $N=1$ to $N=2$ supersymmetry and,
at the same time, for finding a Bogomol'nyi bound \cite{Bogo,dVS}
for the energy of the Abelian Higgs model.

It is interesting to note that Weyl transformations for fermion fields,
in correspondance with (\ref{weyl}) bring, on the one hand the
gravitino Lagrangian to its usual Rarita-Schwinger form. On the
other hand, as we shall see, the Higgs potential and the
Higgs current take after the scaling its usual form as can be seen in the
resulting Bogomol'nyi equations.
Under this Weyl scaling
\beq
\Psi_M \rightarrow \left(-\frac{3}{\kappa^2\ups}\right)^{1/4}\Psi_M \;\;\; ,
\;\;\;
\Lambda \rightarrow \left(-\frac{3}{\kappa^2\ups}\right)^{-3/4}\Lambda \;\;\; ,
\;\;\;
\Xi \rightarrow \left(-\frac{3}{\kappa^2\ups}\right)^{-1/4}\Xi
\label{weylfer}
\eeq
\[ \epsilon \rightarrow \left(-\frac{3}{\kappa^2\ups}\right)^{1/4}\epsilon \]
one has
\[ \delta \Lambda_+\vert = \half F_{MN}\Sigma^{MN}\epsilon_+ +
\frac{i\tilde{q}}{4}(|\phi|^2 - v^2)\epsilon_+ \;\;\;\;\;
, \;\;\;\;\; \delta \Xi_-\vert = \half (\dslash\phi ) \epsilon_+ \]
\beq
\delta {\Psi_M}_-\vert = \frac{2}{\kappa}\left({\cal D}_M + \frac{i\kappa^2}{4}
(J_M + \tilde{q}v^2 A_M)\right) \epsilon_-
\label{susytraf}
\eeq
where
\beq
J_M = \ihalf \left(\phi(D_M\phi)^{*} - \phi^{*}(D_M\phi)\right)
\label{current}
\eeq
is the Higgs field current.

Although we will finally put all fermion fields to zero, we will need for
further calculations the explicit form of the kinetic Lagrangian for fermions.
After the Weyl rescaling (\ref{weylfer}) is performed in Lagrangian
(\ref{lagferkin}), one can note that the kinetic fermionic part can be
diagonalized by the following shift:
\beq
{\Psi_M}_- \rightarrow {\Psi_M}_- - \frac{1}{\kappa\ups}\upc\G_M\Xi_+
\eeq
such that the Fermion Lagrangian can be finally written as:
\bea
V^{-1}{L}_{Fer} & = &
- \frac{1}{2}{\epsilon^{MNRS}}{V}
{\opsim}\Gamma_5\Gamma_N\left({\cal D}_R +
i\frac{\tilde{q}v^2\kappa^2}{4}A_R\G_5\right)\Psi_S \nonumber \\
& - & \half\oxim\left(\not\!\!{\cal D}
- i\left(\tilde{q}+\frac{\tilde{q}v^2\kappa^2}{4}\right)\aslash\right)\Xi
\nonumber\\
& - & \half\ola\left(\not\!\!{\cal D} -
i\frac{\tilde{q}v^2\kappa^2}{4}\aslash\G_5\right)
\Lambda + V^{-1}{\tilde{\cal L}}_{Fer}^{int}
\label{lagferult}
\eea
once the above mentioned conditions for
${\cal J}(\phi,\phi^{*})$ and $\xi$ are adopted.


\section{Dimensional Reduction and Extended supersymmetry}
We shall now derive the $d=3$, $N=2$ Lagrangian
by dimensional reduction of (\ref{lagbosult}).
To this end, we write the vierbein as:
\beq
V_M^A = \left( \begin{array}{cc} e_{\mu}^a & a_{\mu} \\
	0 & \varphi \end{array} \right)
\label{vierbein}
\eeq
(we use $\mu = 0,1,2$ for curved coordinates and $a = 0,1,2$ for flat
indices)
and suppose that whole set of fields
are $x_3$-independent. We will accordingly choose $x_3$ as the
variable which is eliminated within the dimensional reduction procedure.
We introduce in (\ref{vierbein})
$e_{\mu}^a$ as the dreibein of the reduced $3$-dimensional manifold,
$a_{\mu}$ as a vector field and $\varphi$ as a real scalar.
We have chosen
the gauge $V_3^a = 0$, which can always be attained by a suitable
local Lorentz transformation \cite{CJSS}. Indeed,
an infinitesimal field variation of $V_M^A$ under local supersymmetry,
local Lorentz and general coordinate transformations reads
\beq
\delta V_M^A = -i\kappa\opar\G^A\Psi_M + \omega^A_BV_M^B +
\partial_M\eta^RV_R^A + \eta^R\partial_RV_M^A
\label{transfvierbein}
\eeq
where $\opar$, $\omega^A_B$ and $\eta^R$ are the corresponding local
parameters.
Freedom associated with general coordinate invariance in four dimensions
can be exploited to put
\beq
a_{\mu} = 0 \;\;\;\;\; , \;\;\;\;\; \varphi = 1
\label{gauge}
\eeq
without spoiling
the local Einstein group of the reduced $3$-dimensional spacetime.

With these conditions, the metric tensor reads
\beq
g^{MN} = V^M_{A}V^N_{B}\eta^{AB} =
	\left( \begin{array}{cc} g^{\mu\nu} & 0 \\
	0 & - 1 \end{array} \right)
\label{metric}
\eeq
where $\eta^{AB} = diag(+---)$.
Concerning the spinorial connection, it takes the form
\beq
\Omega_{cab}\vert = \omega_{cab}\vert
= - \half (e_c^me_a^n - e_a^me_c^n)\partial_ne_{mb}
		   + \half e_a^me_b^n\partial_ne_{mc} - (a\leftrightarrow b)
\label{spincon}
\eeq
while all other components vanish.
Then, the Ricci scalar in four dimensions ${\cal R}$ reduces to the
corresponding one in $d=3$, which will be denote as $R$.

After dimensional reduction, Lagrangian (\ref{lagbosult}) becomes
\bea
e^{-1}L\vert & = & - \frac{1}{2\kappa^2} R -
\frac{1}{4}g^{\mu\rho}g^{\nu\sigma}F_{\mu\nu}F_{\rho\sigma}
- \half(D_{\mu}\phi)(D^{\mu}\phi)^{*} \nonumber \\
& + & \half\dmu S\partial^{\mu} S
+ \frac{\tilde{q}^2}{2}S^2|\phi|^2
- \frac{\tilde{q}^2}{8}(|\phi|^2 - v^2)^2
\label{lagbosen3}
\eea
where we have identified
\[ A_M \equiv \left(A_{\mu},S\right).  \]
Lagrangian (\ref{lagbosen3}) describes the dynamics of
the Bosonic sector for the $d=3$
Abelian Higgs model coupled to $N=2$ supergravity.
We will now focus on the dimensional reduction of supersymmetric
transformation rules written in eq.(\ref{susytraf}). To this end,
let us specify our representation for the Clifford algebra
\[ \Gamma^a = \gamma^a\otimes\tau_3 \;\;\; , \;\;\;
\Gamma^3 = 1\otimes i\tau_2  \;\;\; , \;\;\;
\Gamma^5 = 1\otimes \tau_1  \]
\beq
\Sigma^{ab} = \sigma^{ab}\otimes 1 \;\;\; , \;\;\;
\Sigma^{a3} = -\Sigma^{3a} = \gamma^a\otimes\tau_1
\label{gammas}
\eeq
where $\gamma^a$ are $2\times 2$ Dirac matrices for $d=3$ and
$\sigma^{ab} = 1/2 [\gamma^a,\gamma^b]$.
In this basis the original $4$-dimensional Majorana spinors take the form
\beq
\Psi = \left(\begin{array}{c} \Psi_{1} \\ i\Psi_2
\end{array} \right)
\eeq
where $\Psi_1$ and $\Psi_2$ are $2$-component real spinors which can be
taken as $3$-dimensional Majorana fields.
Finally, with these Majorana fields one can construct a
Dirac spinor $\psi$ in $3$ dimensions
\[ \psi = \Psi_1 + i\Psi_2. \]

The dimensionally reduced supersymmetric transformations
for fermions read
\[ \delta \lambda\vert = \half F_{\mu\nu}\sigma^{\mu\nu}\epsilon +
\frac{i\tilde{q}}{4}(|\phi|^2 - v^2)\epsilon + \gamma^{\mu}\dmu S\epsilon \]
\beq
\delta \chi\vert = \half \left(\dslash\phi + i\tilde{q}S\phi\right) \epsilon
\;\;\; , \;\;\;
\delta \psi_3\vert = - i\frac{\tilde{q}\kappa}{2}S(|\phi|^2 - v^2)
\epsilon
\label{trafen3}
\eeq
\[ \delta \psi_{\mu}\vert = \frac{2}{\kappa}\left({\cal D}_{\mu} +
i\frac{\kappa^2}{4}(J_{\mu} + \tilde{q}v^2 A_{\mu})\right) \epsilon \equiv
\frac{2}{\kappa}\hat{\nabla}_{\mu}\epsilon.    \]
Note that we have included the transformation for $\psi_3$, a remnant
of the $4$ dimensional starting model. However, as we shall see, $S$ will
be put to zero to recover the Abelian Higgs model in the
bosonic sector, this giving a trivial $\psi_3$ transformation rule.
Being $\epsilon$ a Dirac fermion, transformations (\ref{trafen3}) are $N=2$
supersymmetric ones.
Supersymmetric covariant derivative $\hat{\nabla}_{\mu}$ is defined as
\beq
\hat{\nabla}_{\mu}\epsilon = \left(\partial_{\mu}
+ \frac{1}{4}\omega_{\mu ab}\sigma^{ab} + i\frac{\kappa^2}{4}
(J_{\mu} + \tilde{q}v^2 A_{\mu})\right) \epsilon
\label{supcovder}
\eeq

We shall end this Section by performing the dimensional reduction of the
fermionic Lagrangian (\ref{lagferult}) which is the fermionic counterpart
of the $N=2$ bosonic Lagrangian (\ref{lagbosen3}).
In order to achieve a diagonalized kinetic fermionic sector, we perform
the following shift
\[ \psi_{\mu} \rightarrow \psi_{\mu} + i\gamma_{\mu}\psi_3  \]
after which, the resulting fermionic Lagrangian reads:
\bea
e^{-1}{L}_{Fer} & = &
- \half\frac{\epsilon^{\rho\mu\sigma}}{e}
\overline{\psi}_{\rho}\partial_{\mu}\psi_{\sigma} -
\half\overline{\lambda}\not\!\partial\lambda
- \half\overline{\chi}\not\!\partial\chi \nonumber \\
& - & i\overline{\psi}_3\not\!\partial\psi_3
+ e^{-1}{\hat{L}}_{Fer}^{int}
\label{lagferen3}
\eea
where the last term $\hat{L}_{Fer}^{int}$ includes also gauge interactions.


\section{Self-duality equations and the
Bo\-go\-mol'\-nyi bound}
In this Section we shall obtain a Bogomol'nyi bound for
field configurations in the $d=3$
Abelian Higgs model coupled to gravity. In doing this,
we shall make explicit the relation
between this bound and the supercharge algebra of the $N=2$ theory.

The equations of motion for bosonic matter
fields are:
\beq
\frac{1}{\metric}\partial_{\mu}(\metric F^{\mu\nu}) = - \tilde{q}J^{\nu}
\label{elamu}
\eeq
\beq
\frac{1}{\metric}D_{\mu}(\metric D^{\mu}\phi)
= \frac{\tilde{q}^2}{2}(|\phi|^2 - v^2)\phi
- \tilde{q}^2S^2\phi
\label{elphi}
\eeq
\beq
\frac{1}{\metric}\dmu(\metric\partial^{\mu}S) = \tilde{q}^2|\phi|^2 S.
\eeq
Concerning Einstein equations
\beq
R_{\mu\nu} - \half g_{\mu\nu}R = T_{\mu\nu}^{mat},
\label{einstein}
\eeq
\bea
T_{\mu\nu}^{mat} & = & - g^{\lambda\tau}F_{\mu\tau}F_{\nu\lambda}
- \half(D_{\mu}\phi)(D_{\nu}\phi)^{*}
- \half(D_{\mu}\phi)^{*}(D_{\nu}\phi)
- \partial_{\mu}S\partial_{\nu}S \nonumber \\
& + & g_{\mu\nu}\left[\frac{1}{4}F_{\rho\sigma}F^{\rho\sigma} +
\half(D_{\rho}\phi)(D^{\rho}\phi)^{*} - \half\partial_{\rho}S\partial^{\rho}S
- \frac{\tilde{q}^2}{2}S^2|\phi|^2 \right. \nonumber \\
& + & \left. \frac{\tilde{q}^2}{8}(|\phi|^2 - v^2)^2\right]
\label{tmunu}
\eea

Since we shall focus
on the Abelian Higgs model coupled
to gravity, we make at this point $S = 0$.
Moreover, since Bogomol'nyi equations correspond to static
configurations with $A_0 = 0$, we also impose these conditions
(note that in this case $T_{0i}^{mat} = 0$).

Concerning the metric, let us notice that
a static spacetime admits a surface $\Pi$ orthogonal everywhere to the
time-like killing vector field $\frac{\partial}{\partial t}$.
In a local chart adapted to $\frac{\partial}{\partial t}$,
the line element can be written as:
\beq
ds^2 = dt^2 + g_{ij}dx^idx^j
\label{metric1}
\eeq
where $g_{ij}$ is a function depending only on spatial variables
that span $\Pi$. Now, it is well-known
that any $2$-dimensional metric is K\"{a}hler and then we
write the interval in the form
\beq
ds^2 = dt^2 - \Omega^2dzd\ze
\label{metric2}
\eeq
where $\Omega$ is a function of the conformal coordinates
$\Omega(z,\ze)$. Note that for any finite energy configuration,
Einstein equations (\ref{einstein}) constrain the asymptotic
behaviour of $\Omega$ to be
\beq
\Omega(z,\ze) \rightarrow (z\ze)^{-\frac{\kappa^2M}{2}}
\label{omegaasymp}
\eeq
so that the metric is asymptotic to a flat cone with deficit angle
$\delta = \kappa^2M$ \cite{J}, $M$ being related to the source mass
\bea
M & = & \frac{1}{4\pi} \int dzd\ze \Omega^2 T_{tt}^{mat} =
\frac{1}{4\pi} \int dzd\ze \Omega^2 \left[\frac{1}{2}F_{z\ze}F^{z\ze} \right.
\nonumber \\
& + &
\left. \half(D_{z}\phi)(D^{z}\phi)^{*} + \half(D_{\ze}\phi)(D^{\ze}\phi)^{*}
+ \frac{\tilde{q}^2}{8}(|\phi|^2 - v^2)^2\right].
\label{muno}
\eea

The spacetime metric (\ref{metric2})
can be generated by the following dreibein
\beq
e_t^0 = e_{\ze}^- = 1   \;\;\;\;\; , \;\;\;\;\; e_{z}^+ = \Omega^2
\label{dreiveins}
\eeq
(all the other components vanishing) with the flat metric written in conformal
coordinates as
\[   \eta_{00} = - 2\eta_{+-} = - 2\eta_{-+} = 1.       \]
The non-vanishing component of the spinorial connection is:
\beq
\omega_{\ze +-} = - \overline{\partial}\log\Omega
\label{nonvancon}
\eeq
(here $\partial\equiv\partial_z$ and $\opartial\equiv\partial_{\ze}$).

We shall now analyse the $N=2$ algebra of supercharges
for our model.
To construct these charges we shall follow the Noether method.
The conserved current associated with local supersymmetry is given by:
\beq
{\cal J}^{\mu}[\epsilon] =
\sum_{\Phi} \frac{\delta L}{\delta \nabla_{\mu}\Phi}\delta_{\epsilon}\Phi +
\sum_{\Psi} \frac{\delta L}{\delta \nabla_{\mu}\Psi}\delta_{\epsilon}\Psi -
\theta^{\mu}[\epsilon]
\label{jmu}
\eeq
where $\Phi$ and $\Psi$ represent the whole set of bosonic and fermionic
fields respectively.
Concerning $\theta^{\mu}[\epsilon]$, it is defined
through
\beq
\delta_\epsilon S = \int d^3x \nabla_{\mu}\theta^{\mu}_{\epsilon}.
\label{thetamu}
\eeq
In the present case we have:
\bea
\theta^{\mu}[\epsilon] & = &
-\frac{3}{2\kappa}\frac{\epsilon^{\mu\rho\sigma}}{e}\overline{\psi}_{\rho}
\hat{\nabla}_{\sigma}\epsilon - \half\opar(D_{\mu}\phi)\chi
+ \half\opar\gamma_{\nu}F^{\mu\nu}\lambda
- \frac{1}{8}\overline{\chi}\gamma^{\mu}(\dslash\phi) \nonumber \\
& - & \frac{1}{4}\overline{\lambda}\gamma^{\mu}\left(\half
F_{\mu\nu}\sigma^{\mu\nu}
+ \frac{i\tilde{q}}{4}(|\phi|^2 - v^2)\right)\epsilon + h.c. + \ldots
\eea
where the dots indicate terms containing products of three fermion fields,
which are not relevant for our construction since, after computing the
supercharge algebra, we will put all fermions to zero.
Inserting $\theta^{\mu}$ in (\ref{jmu}) we obtain the following
expression for the supersymmetry
current:
\bea
{\cal J}^{\mu}[\epsilon] & = &
- \overline{\lambda}\gamma^{\mu}\left(\half
F_{\alpha\beta}\sigma^{\alpha\beta} + \frac{i\tilde{q}}{4}(|\phi|^2 -
v^2)\right)\epsilon
- \half\overline{\chi}\gamma^{\mu}(\dslash\phi)\epsilon \nonumber \\
& - & \frac{2}{\kappa}\frac{\epsilon^{\rho\mu\sigma}}{e}\overline{\psi}_{\rho}
\hat{\nabla}_{\sigma}\epsilon + h.c. + \ldots
\label{jmufin}
\eea

The conserved charges associated with (\ref{jmufin}),
\beq
{\cal Q}[\epsilon] = \int_{\Sigma}{\cal J}^{\mu}[\epsilon]d\Sigma_{\mu}
\equiv {\cal Q}_1[\epsilon] + i {\cal Q}_2[\epsilon],
\label{qjmu}
\eeq
are defined over a space-like surface $\Sigma$ whose area element is
$d\Sigma_{\mu}$.
Here ${\cal Q}_I$, {\scriptsize I=1,2}, are the Majorana charge generators.

Now, imposing the
gravitino field equation,
\beq
\frac{2}{\kappa}\frac{\epsilon^{\mu\sigma\rho}}{e}
\overline{\hat{\nabla}_{\sigma}\psi}_{\rho} =
\overline{\lambda}\gamma^{\mu}\left(\half
F_{\mu\nu}\sigma^{\mu\nu} + \frac{i\tilde{q}}{4}(|\phi|^2 - v^2)\right)
+ \half\overline{\chi}\gamma^{\mu}(\dslash\phi)
\eeq
one can see from eqs.(\ref{jmufin})-(\ref{qjmu}),
after integration by parts,
that the supercharge is nothing but
the circulation of the gravitino
arround the oriented boundary $\partial\Sigma$
\beq
{\cal Q}[\epsilon] = -
\frac{2}{\kappa}\oint_{\partial\Sigma}\opar\psi_{\mu}dx^{\mu}
\label{qsurf}
\eeq
Let us stress that the expression above coincides with the results
presented by Teitelboim
\cite{T} for pure supergravity in $4$-dimensional spacetime,
after dimensional reduction. Although there are well-known problems
for constructing supergravity charges in $2 + 1$ dimensions
\cite{W1}-\cite{W2},
we shall see that they can be overcome in the present model.

As explained in \cite{T},
it is not possible to compute the supercharge algebra by (naively) evaluating
Posson brackets from
(\ref{qsurf}) because surface terms do not have
well defined functional derivatives and hence their Poisson brackets with
the various fields of the theory are not well defined. One can compute
instead the algebra by acting on the integrand
of (\ref{qsurf}) with a supersymmetry transformation:
\beq
\{\bar{{\cal Q}}[\epsilon],{\cal Q}[\epsilon]\}\vert \equiv
\delta_{\epsilon}{\cal Q}[\epsilon]\vert =
\frac{2}{\kappa}\oint_{\partial\Sigma}\opar\delta_{\epsilon}\psi_{\mu}dx^{\mu}
=
\frac{4}{\kappa^2}\oint_{\partial\Sigma}\opar\hat{\nabla}_{\mu}\epsilon
dx^{\mu}
\label{qeqe}
\eeq
where $\bar{{\cal Q}}$ is given by:
\beq
\bar{{\cal Q}} =
\frac{2}{\kappa}\oint_{\partial\Sigma}\overline{\psi}_{\mu}\epsilon dx^{\mu}
\label{qraya}
\eeq

Now, Teitelboim \cite{T} has proven, using Dirac formalism for constrained
systems, that supergravity
charges, which can be written as surface integrals in the form (\ref{qsurf}),
obey a flat-space
supersymmetry algebra
\beq
\{\bar{{\cal Q}}_I,{\cal Q}_J\} =
\delta_{IJ}\gamma^{\mu}P_{\mu} + \epsilon_{IJ}Z
\label{global}
\eeq
where $Z$ is the central charge.
In flat space, it is a well-known result
that this algebra leads in several models to
Bogomol'nyi bounds for the energy \cite{WO}-\cite{ENS}.
Indeed, squaring eq.(\ref{global}) and tracing over the indices, one
obtains a bound
\[   P^{2} - Z^{2} \geq 0 \]
from which, after using the identity of $Z$ with the topological
charge of the field
configuration $T$ \cite{WO}-\cite{ENS},
one obtains the well-known Bogomol'nyi bound
for the mass $M$ of the configuration
\beq
M \geq |T|
\label{bou}
\eeq

Coming back to supergravity,
let us see explicitely how (\ref{qeqe}) ensures that static
finite-energy configurations satisfy a bound of topological nature
of the type (\ref{bou}).
To this end, let us write, using the expression
for the covariant derivative given in (\ref{supcovder})
\bea
\{\bar{{\cal Q}}[\epsilon],{\cal Q}[\epsilon]\}\vert & = &
\frac{4}{\kappa^2}
\oint_{\partial\Sigma}\opar\hat{\nabla}_{\mu}\epsilon dx^{\mu} \nonumber \\
& = & \frac{4}{\kappa^2}\oint_{\partial\Sigma}\opar{\cal D}_{\mu}\epsilon
dx^{\mu} + i\oint_{\partial\Sigma}\opar\epsilon(J_{\mu} +
\tilde{q}v^2A_{\mu}) dx^{\mu}
\label{qeqeuno}
\eea
We can now use the asymptotic behaviour of different fields appearing in
(\ref{qeqeuno}). The spinorial connection which enters in the covariant
derivative in the first term of the r.h.s. behaves as
\beq
\omega_{\ze +-} \rightarrow \frac{\kappa^2M}{2\ze}
\label{omeasy}
\eeq
Concerning the electromagnetic field as well as the Higgs current,
finiteness of the energy implies
\beq
A_{z} \rightarrow - \frac{in}{\tilde{q}z} \;\;\; , \;\;\;
A_{\ze} \rightarrow \frac{in}{\tilde{q}\ze} \;\;\; , \;\;\;
J_{z} \rightarrow O\left(\frac{1}{z\ze}\right) \;\;\; , \;\;\;
J_{\ze} \rightarrow O\left(\frac{1}{z\ze}\right).
\label{ajasy}
\eeq
(Here,
the integer $n$ is the topological number that characterizes the quanta of
magnetic flux).
Finally, the asymptotic behaviour of $\epsilon$ will be written in the
form:
\beq
\epsilon\rightarrow\Theta(z\ze)\epsilon_{\infty}
\label{compasimp}
\eeq
where $\Theta(z\ze)$ will be determined using the so-called
Witten condition\cite{W} (see below).

We can now evaluate the line-integral (\ref{qeqeuno}). To avoid
infrared divergences, it is necessary to take the contour of integration
at large but finite radius $R$.
Using the asymptotic behaviours listed above, (\ref{qeqeuno}) becomes:
\beq
\{\bar{{\cal Q}}[\epsilon],{\cal Q}[\epsilon]\}\vert =
(M\opar_{\infty}\gamma^0\epsilon_{\infty}
- v^2n\opar_{\infty}\epsilon_{\infty})\Theta(R)^{2}.
\label{lns}
\eeq
The relation of this result with the Poisson brackets (\ref{global})
which are valid, in particular, in flat space, is, of course, no accident:
as we discussed above,
supersymmetry transformations at spatial infinity generate global
$N=2$ supersymmetry algebra.

We can now prove that
$\{\bar{{\cal Q}}[\epsilon],{\cal Q}[\epsilon]\}\vert$
is semi-positive definite and then derive a Bogomol'nyi
bound of topological origin from (\ref{lns}).

First, let us observe that the supercharge algebra
evaluated in the purely bosonic sector
is the integral over the boundary of a $1$-form $\omega$,
constructed
from a fermionic parameter:
\beq
\{\bar{{\cal Q}[\epsilon]},{\cal Q}[\epsilon]\}\vert =
\frac{4}{\kappa^2}\oint_{\partial\Sigma} \omega
\label{qeqeult}
\eeq
\beq
\omega = \opar\hat{\nabla}_{\mu}\epsilon dx^{\mu}
\label{nester}
\eeq
Now, $\omega$ can be identified with the generalized
Nester-like form \cite{W,NI} in 3 dimensional spacetime.
The use of Nester form is at the root of several proofs
 of the positivity of gravitational energy
\cite{T}-\cite{NI} and it was also used in $4$ and
$5$-dimensional models to find Bogomol'nyi bounds
\cite{Gibb}-\cite{CGR}.
In the same vein,
we shall see below that the integral of $\omega$ on the contour is
semi-positive definite and that, as a consequence, the theory posses
a Bogomol'nyi bound. Let us mention at this point that the integral in the
r.h.s. of (\ref{qeqeult}) coincides with
the quantity $\Delta (r)$ introduced in Ref.\cite{BBS}. Our construction
shows that its occurence is a consequence of the $N=2$ supercharge algebra.

First, using Stokes' theorem, we have
\beq
\oint_{\partial\Sigma}\omega =
\int_{\Sigma}\epsilon^{\mu\nu\beta}\hat{\nabla}_{\beta}
(\opar\hat{\nabla}_{\mu}\epsilon)d\Sigma_{\nu}.
\label{cadorna}
\eeq
where the integrand in (\ref{cadorna}) can be written as
\beq
\epsilon^{\mu\nu\beta}\hat{\nabla}_{\beta}
(\opar\hat{\nabla}_{\mu}\epsilon) = \epsilon^{\mu\nu\beta}
\overline{\hat{\nabla}_{\beta}\epsilon}\hat{\nabla}_{\mu}\epsilon
+ \half\epsilon^{\mu\nu\beta}\overline\epsilon
[\hat{\nabla}_{\beta},\hat{\nabla}_{\mu}]\epsilon ~.
\label{divnester}
\eeq
Then, using
\beq
[\hat{\nabla}_{\mu},\hat{\nabla}_{\nu}] = \half {R_{\mu\nu}}^{ab}\Sigma_{ab}
+ \frac{i\tilde{q}v^2\kappa^2}{2}F_{\mu\nu} +
\frac{i\kappa^2}{2}(\partial_{\mu}J_{\nu} -
\partial_{\nu}J_{\mu}),
\label{nmunu}
\eeq
Einstein
equations (\ref{einstein})
and supersymmetric transformations for the
fermionic fields $\lambda$ and $\chi$,
we arrive
to the following expression
\beq
\epsilon^{\mu\nu\beta}\hat{\nabla}_{\beta}
(\opar\hat{\nabla}_{\mu}\epsilon) = \epsilon^{\mu\nu\rho}
\overline{\hat{\nabla}_{\mu}\epsilon}\hat{\nabla}_{\rho}\epsilon
+ \frac{\kappa^2}{2}\left[
\overline{\delta_\epsilon\lambda}\gamma^{\nu}\delta_\epsilon\lambda
+ \overline{\delta_\epsilon\chi}\gamma^{\nu}\delta_\epsilon\chi \right] ~.
\label{divernest}
\eeq
We now specialize our spacelike integration
surface $\Sigma$ so that $d\Sigma_{\mu} =
(d\Sigma_{t},\vec{0})$. Then, we only need to compute the time component
of eq.(\ref{divernest}) which, after some Dirac algebra, reads
\bea
\epsilon^{t\nu\beta}\hat{\nabla}_{\beta}
(\opar\hat{\nabla}_{\mu}\epsilon) & = &
\left(\gamma^i\hat{\nabla}_i\epsilon\right)^{\dag}
\left(\gamma^j\hat{\nabla}_j\epsilon\right) -
g^{ij}\left(\hat{\nabla}_i\epsilon\right)^{\dag}
\left(\hat{\nabla}_j\epsilon\right) \nonumber \\
& + &
\frac{\kappa^2}{2}
\left[{\delta_{\epsilon}\lambda}^{\dag}\delta_{\epsilon}\lambda
+ {\delta_{\epsilon}\chi}^{\dag}\delta_{\epsilon}\chi \right]
\label{compcero}
\eea
We see at this point, that if we impose the
generalized Witten condition \cite{W}
\beq
\gamma^i\hat{\nabla}_i\epsilon = 0
\label{genw}
\eeq
the r.h.s. of eq.(\ref{compcero}) is semi-positive definite
\beq
\epsilon^{t\nu\beta}\hat{\nabla}_{\beta}
(\opar\hat{\nabla}_{\mu}\epsilon) \geq 0
\label{cota}
\eeq
and then, using
(\ref{cadorna}),
\beq
\{\bar{{\cal Q}}[\epsilon],{\cal Q}[\epsilon]\}\vert \geq 0
\label{supbound}
\eeq
The bound is saturated if and only if
\beq
\delta_{\epsilon}\lambda = 0
\label{deluno}
\eeq
\beq
\delta_{\epsilon}\chi = 0
\label{deldos}
\eeq
\beq
\hat{\nabla}_{i}\epsilon = 0
\label{deltacero}
\eeq

Condition (\ref{deltacero}) reflects our choice of surface $\Sigma$ .
One can easily see that a
more general choice would imply, instead of (\ref{deltacero}),
\beq
\hat{\nabla}_{\mu}\epsilon = 0 ~.
\label{ndeltacero}
\eeq

In explicit form, eqs.(\ref{deluno}), (\ref{deldos}) and (\ref{ndeltacero})
read:
\beq
\delta_{\epsilon}\lambda = \half \left[F_{\mu\nu}\sigma^{\mu\nu} +
\frac{i\tilde{q}}{2}(|\phi|^2 - v^2)\right]\epsilon = 0
\label{bogo1}
\eeq
\beq
\delta_{\epsilon}\chi = \half (\dslash\phi)\epsilon = 0
\label{bogo2}
\eeq
\beq
\delta_{\epsilon}\psi_{\mu} = \left({\cal D}_{\mu} + i\frac{\kappa^2}{4}
(J_{\mu} + \tilde{q}v^2 A_{\mu})\right) \epsilon = 0
\label{bogo3}
\eeq
We can see at this point that solutions of (\ref{bogo1}) and
(\ref{bogo2}) break half of the supersymmetry. Indeed, writting
\beq
\epsilon \equiv \left( \begin{array}{c} \epsilon_+ \\ \epsilon_- \end{array}
\right)
\label{quiral}
\eeq
One can easily see that the conditions
\beq
\delta_{\epsilon_+}\lambda = 0
\label{deluno+}
\eeq
\beq
\delta_{\epsilon_+}\chi = 0
\label{deldos+}
\eeq
imply $\delta_{\epsilon_-}\lambda \neq 0$,
$ \delta_{\epsilon_-}\chi \neq 0$
for nontrivial solutions.
Hence if one is to search Bogomol'nyi equations for non-trivial
configurations, it
makes sense to consider that $\epsilon$ has just one independent
complex component,
\beq
\epsilon \equiv \left( \begin{array}{c} \epsilon_+ \\ 0 \end{array} \right)
\label{newquiral}
\eeq
satisfying equation (\ref{genw}), which then reads
\beq
\hat{\nabla}_{z}\epsilon_+ = 0
\label{witconfin}
\eeq

Let us stress on the fact that field configuration solving
Bogomol'nyi equations (which as we shall see coincide with (\ref{deluno+})
and (\ref{deldos+})) break half of the supersymmetries,
a common feature in all models presenting
Bogomol'nyi bounds with supersymmetric extension (See for example
\cite{hull,KLOPvP}).
It can be understood as follows:
The number
of Killing spinors (those that solve eq.(\ref{bogo3}))
admitted by a certain spacetime coincides with
the number of remnant unbroken supersymmetries \cite{hull}.
If we attempt to keep all the supersymmetries of our model,
we will find that the resulting field configuration has zero energy
(the trivial vacuum) as it was remarked above.
Then, it is evident that we must restrict the space
of solutions of (\ref{bogo3}) if we want to obtain non-trivial topological
configurations (which, in this sense, require breaking of one of the
supersymmetries).

We can now study the asymptotic behaviour of the Killing spinor
$\epsilon_+\rightarrow\Theta(R)\epsilon_{+\infty}$.
One can see that eq.(\ref{witconfin}) implies that $\Theta(R)$ is
a power of the radius $R$:
\beq
\Theta(R) = R^{-\kappa^2\frac{nv^2}{2}}
\label{theta}
\eeq
It is important, at this point,
to comment on the existence of non-trivial solutions to eq.(\ref{witconfin}) in
asymptotically conical spaces. In fact, the supercovariant derivative
in (\ref{witconfin}) gets an electromagnetic contribution
related to the magnetic flux,
and then, as explained in \cite{BBS}, leads
to the existence of non-trivial solutions, otherwise absent \cite{W}.

Now, for a parameter $\epsilon$ of the form (\ref{newquiral}),
formula (\ref{lns}) can be written as
\beq
\{\bar{{\cal Q}}[\epsilon],{\cal Q}[\epsilon]\}\vert =
(M - v^2n)\epsilon^{\dag}_{+\infty}\epsilon_{+\infty}\Theta(R)^{2}
\label{newlns}
\eeq

We can now finally write a Bogomol'nyi
bound using (\ref{supbound}) and (\ref{newlns}):
\beq
M - nv^2 \geq 0
\label{bog}
\eeq

The mass of our field configuration, defined in eq.(\ref{muno}),
is then bounded from below
by the magnetic flux quanta. Analogous results also hold in the
$4$ and $5$ dimensional models studied by Kallosh et al \cite{KLOPvP} and
Gibbons et al \cite{GKLTT} where the
bound reads: $M\geq\sqrt{Q^2 + P^2}$, $Q$ and $P$ being the
electric and magnetic charges respectively, related
to the two central charges existing in the extended supersymmetry
algebra of those models \cite{hull}. In the present
abelian $3$-dimensional model there is just one central charge and,
moreover, it is not possible to have electrically
charged finite energy solitonic configurations in the Abelian Higgs model
so that $Q = 0$.
\vspace{3mm}

\noindent
The bound (\ref{bog}) for the $d=3$ Abelian Higgs model coupled to
gravity coincides with that presented in \cite{BBS}.
The novelty here is that we have obtained it from the supercharge
algebra (plus the generalized Witten condition)
thus showing that the connection between global supersymmetry and
Bogomol'nyi bounds discussed in \cite{WO}-\cite{ENS},
also works for \underline{local} supersymmetry.
\vspace{3mm}

The bound (\ref{bog}) is saturated whenever eqs.(\ref{bogo1})-(\ref{bogo3})
hold and hence, we identify them as the Bogomol'nyi equations of our model.
Concerning equations (\ref{bogo1}) and (\ref{bogo2}), they can be written in
the usual form \cite{Bogo}:
\beq
\epsilon^{z\ze}F_{z\ze} +
\frac{\tilde{q}}{2}(|\phi|^2 - v^2) = 0
\label{hoybogo1}
\eeq
\beq
D_z\phi = 0.
\label{hoybogo2}
\eeq
They have been found previously in the study of the Einstein-Abelian Higgs
model by
Comtet and Gibbons in ref.\cite{CG} where it is shown
that this set of equations admits
vortex solutions in $3+1$ dimensions which can be interpreted as cosmic
strings.
Concerning eq.(\ref{bogo3}),
\beq
\left({\cal D}_{\mu} + i\frac{\kappa^2}{4}
(J_{\mu} + \tilde{q}v^2 A_{\mu})\right) \epsilon_+ = 0,
\label{newbogo}
\eeq
it can be thought as the Bogomol'nyi equation
for the gravitational field. Indeed,
it can be seen that the integrability conditions of this equation,
\beq
\left[{\cal D}_{\mu} + i\frac{\kappa^2}{4}(J_{\mu} + \tilde{q}v^2 A_{\mu}),
{\cal D}_{\nu} + i\frac{\kappa^2}{4}(J_{\nu} + \tilde{q}v^2 A_{\nu})\right]
\epsilon_+ = 0,
\label{integra}
\eeq
become the Einstein equations once eqs.(\ref{hoybogo1})-(\ref{hoybogo2})
are imposed.
Even though conical spaces usually does not
admit covariantly constant spinors, solutions to eq.(\ref{newbogo})
exist as a consequence of the electromagnetic contribution to the
supercovariant derivative, as explained above.
In view of the bound (\ref{bog}), the solutions of the whole
set of Bogomol'nyi equations satisfy the more involved second
order Euler-Lagrange ones.

\section{Summary and discussion}

We have studied Bogomol'nyi bounds for the Abelian Higgs model
coupled to gravity in $2+1$ dimensions by embedding it in an extended
$N=2$ supergravity model.

Bogomol'nyi  equations for the scalar and the gauge field,
compatible with the Einstein equation, were originally derived
by Comtet and Gibbons \cite{CG}, following the usual Bogomol'nyi
approach \cite{Bogo}. In their work,
the existence of gravitating
multi-vortex solutions saturating the Bogomol'nyi bound
was proven.
These solutions can be understood as cosmic strings and the bound can be
seen to be a restriction in possible values for the deficit angle of
conical spacetime \cite{CG}.

More recently, Becker, Becker and Strominger \cite{BBS} considered the
model we discussed and obtained the Bogomol'nyi bound
from supersymmetry arguments. They followed an
approach which is a variant of the methods leading to energy positivity
in gravity models \cite{W}-\cite{GH}. This approach is close to the one
we have followed in the present paper. The novelty in our
work is that we have obtained the Bogomol'nyi bound and the self-duality
equations (including the one associated with the Einstein equation)
by constructing the supercharge algebra and making explicit its relation
with the (generalized) Witten-Nester-Israel form.

As it is well-known, in $2+1$ dimensions, states with non-zero energy produce
asymptotically conical spaces and this put stringent limitations on
the existence of covariantly constant spinors which are basic in
the construction of supercharges. However, in our model, the supercovariant
derivative gets a  contribution related to the topological
charge of the vortex and this allows for non-trivial solutions for Killing
spinors. This is the reason why we were able to find supercharges and
determine their algebra.

A possible interest of our investigation is related to the recent discussion
about how can supersymmetry ensure the vanishing of the cosmological
constant \cite{W1},\cite{BBS},\cite{W2}.
On the other side, since our approach provides a systematic way
of investigating Bogomol'nyi bounds in supergravity models in arbitrary
dimensions (as it is the case for global supersymmetry \cite{WO}-\cite{ENS}),
one can think in finding Bogomol'nyi bounds for other
gauge models coupled to gravity like, for example, the Chern-Simons-Higgs model
recently considered in \cite{L}. We hope to come back to these problems in a
forthcoming work.


\end{document}